\documentclass{ws-p8-50x6-00}

\begin{document}

\title{Final state interaction in inclusive and exclusive quasi-elastic
 processes}

\author{M. Alvioli$^1$, M.A. Braun $^2$, C. Ciofi degli Atti$^1$, L.P.
      Kaptari$^3$,\\ H. Morita$^4$, D. Treleani$^5$}
\address{{\small(The Dubna-Hokkaido-St.Petersburg-Perugia-Trieste
collaboration)}}

\address{$^1$ Department of Physics, University of Perugia, and Istituto Nazionale
di Fisica Nucleare, Sezione di Perugia, Via A. Pascoli, I--06100 Perugia,
Italy }

\address{$^2$ Department of High-Energy Physics,
S.Petersburg University, 198904 S.Petersburg, Russia}

\address{$^3$ Bogoliubov Laboratory of  Theoretical  Physics, Joint Institute for
Nuclear Research, Dubna, Russia}
\address{$^4$ Sapporo Gakuin University, Bunkyo-dai 11, Ebetsu 069, Hokkaido, Japan}
\address{$^5$ Department of
Theoretical Physics, University of Trieste, Strada
Costiera 11, Istituto Nazionale di Fisica Nucleare, Sezione di Trieste,
and ICTP, I--34014, Trieste,Italy}

\def\beq{\begin{equation}}
\def\eeq{\end{equation}}
\def\noi{\noindent}

\maketitle

\abstracts{
We discuss
a new approach to final state interactions, that keeps explicitly into account
the virtuality of the
ejected nucleon in
quasi-elastic $A(e,e'p)X$ scattering at very large $Q^2$,
and we present some
recent results, at moderately large $Q^2$ values, for the nuclear transparency
in $^4He$, $^{16}O$ and
$^{40}Ca$ and for the momentum
distributions of $^4He$.}

\section{Finite formation time in quasi-elastic $A(e,e')X$ processes}

Final state interactions in inclusive
quasi-elastic   $A(e,e')X$ processes at large $Q^2$ are
characterized by the large virtuality of
the ejected nucleon. A straightforward way to incorporate virtuality effects in the
process is through the
Feynman diagrams formalism[1].
 The amplitude describing $n$ consecutive rescattering of
the ejectile
emerging from the interaction of the struck nucleon  with the
incoming virtual photon, is represented in the diagram in Fig.1.
\begin{figure}
\vspace{-6 cm}
\centerline{
\epsfysize=19cm \epsfbox{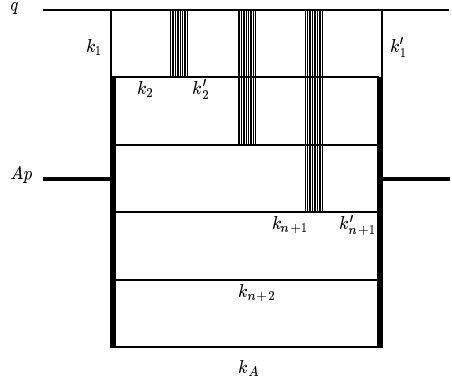}
}
\vspace{-8 cm}
\caption[ ]{The forward scattering amplitude.}
\label{ampl}
\end{figure}
The Glauber expression of the FSI may be obtained from the amplitude of the
process in momentum space, obtained by
the diagram, by applying the following procedure: one should first take
the residues of the target nucleons
propagators, to make the energy integrals of the loop variables. Then,
while taking the Fourier transform to go to the coordinates representation,
one should make the further
approximations of $i$) disregarding the propagators of the
ejected nucleon in the transverse momentum integrals,
and $ii$) of keeping into account only the propagators of the ejected
nucleon when
making the longitudinal momentum integrations.
As a result,  the amplitude is expressed
as an integral over the transverse coordinate $b$ of the struck nucleon,
with respect to the direction of the virtual photon,
and on the longitudinal coordinates of the
interaction
points $z_i$, $i=1,2,...n+2$, with the appropriate nuclear
density matrices.

To introduce the dependence on the virtualities $v_j=(k_1^{(j)})^2-m^2$,
the simplest assumption
on the nucleon-nucleon amplitude
$f_j$, is to use the
factorized form[2]
\beq
f_j=F(v_{j-1})F(v_j)f;\    \ \gamma(k_1, q)=F(v_1)\gamma(Q^2)
\eeq
where $f$ is the on-shell amplitude and $F(v)$  a form-factor
 exhibiting the dependence of $f$ on the virtuality of
the external lines,
normalized according to $F(0)=1$ and decreasing with $v$. It is natural to
introduce the dependence of the off-mass-shell electric
form-factor $\gamma$ on the virtuality of the struck nucleon in the same manner.
Going to the coordinates representation, using the recipe described above, one
obtains in this way that
all integrations on the longitudinal momenta become
factorized. Each term in the product has the form
\beq
iJ(-z)=\int\frac{dv}{2\pi}\frac{F^2(v)}{-v-i0}\exp\left(i\frac{xm}{Q^2}vz\right)
\eeq
so that, if the dependence on the virtuality of the nucleon-nucleon
interaction amplitude is disregarded (namely for $F(v)=1$), one obtains
$J(z)\to\theta(z)$.
The introduction of  the dependence of the amplitudes
on the virtualities,  by means of the
factorized expression (1), is therefore effectively equivalent to the
replacement of  the usual
$\theta(z)$
in the nucleon
propagator with the function
$J(z)$, which depends on the virtuality through the  form-factor $F(v)$.

On rather general grounds one may express the form-factor
squared as
\beq
F^2(v)=\int_0^{+\infty}\frac {dv'v'\tau(v')}{v'-v-i0};\    \ \int_0^{+\infty}
dv\tau(v)=1
\eeq
so that one may write
\beq
J(z)=\theta(z)\int_0^{+\infty}
dv\tau(v)\left(1-\exp\left(-i\frac{xmvz}{Q^2}\right)\right)
\eeq
and, with the simplest choice
$\tau(v)=\delta(v-M^2)$, one obtains
\beq
J(z)=\theta(z)
\Big[1-\exp\left(-i\frac{z}{l(Q^2)}\right)\Big]
\eeq
where
$l(Q^2)=Q^2/(xmM^2)$
has the obvious meaning of a formation length
growing linearly with $Q^2$.

At the level of a single rescattering,
the picture of FSI obtained in this way
 coincides with a standard two-channel Glauber model
for the propagating nucleon and its excited state of mass squared
${m^*}^2=m^2+M^2$.
Indeed the amplitude with a single rescattering is written as
\beq
{\cal A}^{(1)}=
\langle\gamma^2f[1-e^{-i(z'_1-z)/l}-
e^{-i(z-z_1)/l}+e^{-i(z'_1-z_1)/l}]
\rangle_1
\eeq
On the other hand, the two-channel Glauber model with two
ejectile states 1 (the nucleon) and
2 (its excited state) leads to the single rescattering contribution
\beq
{\cal A}^{(1)}=
\langle\gamma_1^2f_{11}+\gamma_1\gamma_2f_{21}e^{-i(z'_1-z)/l}+
\gamma_1\gamma_2f_{12}e^{-i(z-z_1)/l}+
\gamma_2^2f_{22}e^{-i(z'_1-z_1)/l}
\rangle_1
\eeq
where the average $\langle...\rangle$ is defined as
\bea
\langle{\cal O}(b_1, z, z_1, z'_1)\rangle_1\equiv-&&\frac{x^2m}{2}A(A-1)
\int d^2b_1dzdz_1dz'_1{\cal O}(b_1, z, z_1, z'_1)\cr\nonumber
&&\times\rho(b_1,z)\rho(b_1z_1|b_1z'_1)\theta(z'_1-z)\theta(z-z_1)
e^{i\Delta(z'_1-z_1)}
\eea
with $\Delta=Q^2(1-x)/(2q_zx)$, $\rho$ the density matrix,
$f_{ik}=f_{ki}$, $i,k=1,2$ the forward scattering amplitudes for
transitions $i\rightarrow k$ and $\gamma _i$, $i=1,2$ the vertices for the
production of the two ejectile states.  One immediately observes that
(6) and (7) coincide if
\beq
f_{11}\gamma_1+f_{12}\gamma_2=0,\ \
f_{21}\gamma_1+f_{22}\gamma_2=0
\eeq
and, moreover,  if  $\gamma_{11}^2f_{11}$ in (7) is identified
 with $\gamma^2f$ in (6).
The meaning of the sum rules (8)
is that when applying the matrix $f_{ik}$ to the vector $\gamma_i$ one obtains
zero, which, as
discussed
in ref.[3], is precisely the condition for color transparency.
In the case of two channels one may easily see that both unitarity,
$2{\Im}{\rm m}f_{il}=\sum_{j=1,2}f_{ij}f^*_{jl}$, and the transparency
conditions are
satisfied
by
\beq
f_{12}=f_{21}=-\xi f_{11},\ \
f_{22}=\xi^2f_{11}
\eeq
where $\xi$ is the (real) ratio of the form factors $\gamma_1$ and $\gamma_2$,
whose value is obtained by
$\xi^2=|f_{12}|^2/|f_{11}|^2=\sigma_{inel}/\sigma_{el}$. All parameters are
then
fixed by the value of the total and of the elastic nucleon-nucleon cross
sections, namely by
the imaginary part and by the modulus of $f_{11}$.

\section{The deuteron target}

In the simplest case of the deuteron target the expression of the
the amplitude  with a single rescattering is
\bea
{\cal A}^{(1)}&=&-(1/2)\gamma^2x^2m
\int dz_1dz'_1d^2b\psi(b,z_1)i\Gamma(b)\psi(b,z'_1)J(-z_1)J(z'_1)
{\large \sl e}^{i\Delta(z'_1-z_1)}\cr
&=&(1/2)\gamma^2x^2m\int d^2b i\Gamma(b)\bigl[X(b,x,Q^2)\bigr]^2
\eea
where $b$ is the distance
between the proton and the neutron in transverse space,
$\psi(b,z)$ is the deuteron wave function and $\Gamma(b)$ the nucleon-nucleon
profile function.
To obtain the various contributions to the structure function one needs to
consider the different discontinuities of the amplitude. To that purpose
one may write:
\begin{eqnarray}
X(b,x,Q^2)&=&i\int dz\psi(b,z)J(-z)\exp(i\Delta z)\nonumber\\
&=&\int\frac{d^3k}{(2\pi)^{3/2}}\phi(k)e^{i{\bf k}_t{\bf
b}}\Biggl[\frac{1}{k_z-\Delta-i0}-\frac{1}{k_z-\Delta+\frac{1}{l}-i0}\Biggr]
\end{eqnarray}
where $\phi(k)$ is the deuteron  wave function in  momentum space.
The cross-section
to produce a fast nucleon has two different contributions:
from the cut of the amplitude $\Gamma$ and from the cut of the nucleon
propagators.
The sum of the two discontinuities from the cut nucleon propagators
gives
\beq
{\rm Disc}_1{\cal A}^{(1)}=ix^2m\gamma^2\int
d^2bY(b,x){\Re}{\rm e}\bigl[i\Gamma(b)X(b,x,Q^2)\bigr]
\eeq
where
\beq Y(b,x)=
\int\frac{d^3k}{(2\pi)^{3/2}}\phi(k)e^{i{\bf k}_t{\bf
b}}2\pi\delta(k_z-\Delta)
\eeq
and, in the Bjorken limit, is independent on  $Q^2$.
As for the discontinuity corresponding to a cut across the
rescattering
blob $\Gamma$, the contribution of the scattered
nucleon to the inclusive structure function is obtained by taking
only the
elastic part of the unitarity sum over the intermediate states:
\beq
{\rm Disc}_2{\cal A}^{(1)}=i(1/2)x^2m\gamma^2\int
d^2b|\Gamma(b)|^2|X(b,x,Q^2)|^2
\eeq
The  contribution to the inclusive  deuteron structure function due to the
nucleon rescattering in the final state, $F_{2}^{N/d,(1)}$, is given by the
sum of these two
discontinuities divided by $i$.
At low energy, when no elastic channel is open and
$\sigma_{tot}=\sigma_{el}$, one obtains
\begin{eqnarray}
F_{2}^{N/d,(1)}&=&x^2m\gamma^2\int
d^2b\Biggl\{-2{\Im}{\rm
m}\Gamma(b)+\frac{|\Gamma(b)|^2}{2}\Biggr\}\frac{\bigl[Y(b,x)\bigr]^2}{4}
\nonumber\\
&=&x^2m\gamma^2\int d^2b\Biggl\{-{\Im}{\rm m}\Gamma(b)\Biggr\}\frac{\bigl[Y(b,x)
\bigr]^2}{4}=2
{\Im}{\rm m}{\cal A}^{(1)}
\end{eqnarray}
so that the only contributions to the imaginary part of the forward amplitude is
given
by the the two discontinuities (12) and (14), where only the elastic
intermediate state is present.
At higher energies the inelastic
channels become important. The effect is to add further
contributions to the imaginary part of the forward amplitude. As it may be seen
by looking at the behavior of $X$, Eq.(11), as a function of the formation
length
$l$. The additional
contributions give a small correction at low $Q^2$ (small $l$) while they tend
to cancel completely the elastic contribution at large $Q^2$ (large $l$).

For a quantitative evaluation we write the deuteron structure function as
\begin{equation}
F_{2}^d=F_{2}^{N/d,(0)}+F_{2}^{N/d,(1)}+F_{2}^{N^*/d,(1)}
\label{effe2}
\end{equation}
where $F_{2}^{N/d,(0)}$ is the expression in impulse
 approximation, the
correction induced by FSI with a
proton in the final state is $F_{2}^{N/d,(1)}$, and all other contributions
to FSI are represented by $F_{2}^{N^*/d,(1)}$.
In figure 2 we have plotted, as a function of $Q^2$, the ratios
\beq
R_N(Q^2)= 1 +
\left(\frac{F_{2}^{N/d,(1)}(x,Q^2)}
{F_{2}^{N/d,(0)}(x,Q^2)}\right)_{x=1}
\eeq
and
\beq
R^{tot}(Q^2)=1 + \left(\frac{F_{2}^{N/d,(1)}(x,Q^2)+F_{2}^{N^*/d,(1)}(x,Q^2)}
{F_{2}^{N/d,(0)}(x,Q^2)}\right)_{x=1}
\eeq
at $x=1$, for the values $m^*=1.4$, $1.8$, $2.4$ GeV and for the pure Glauber
case.

\begin{figure}
%\vspace{2 cm}
\centerline{
\epsfysize=6cm \epsfbox{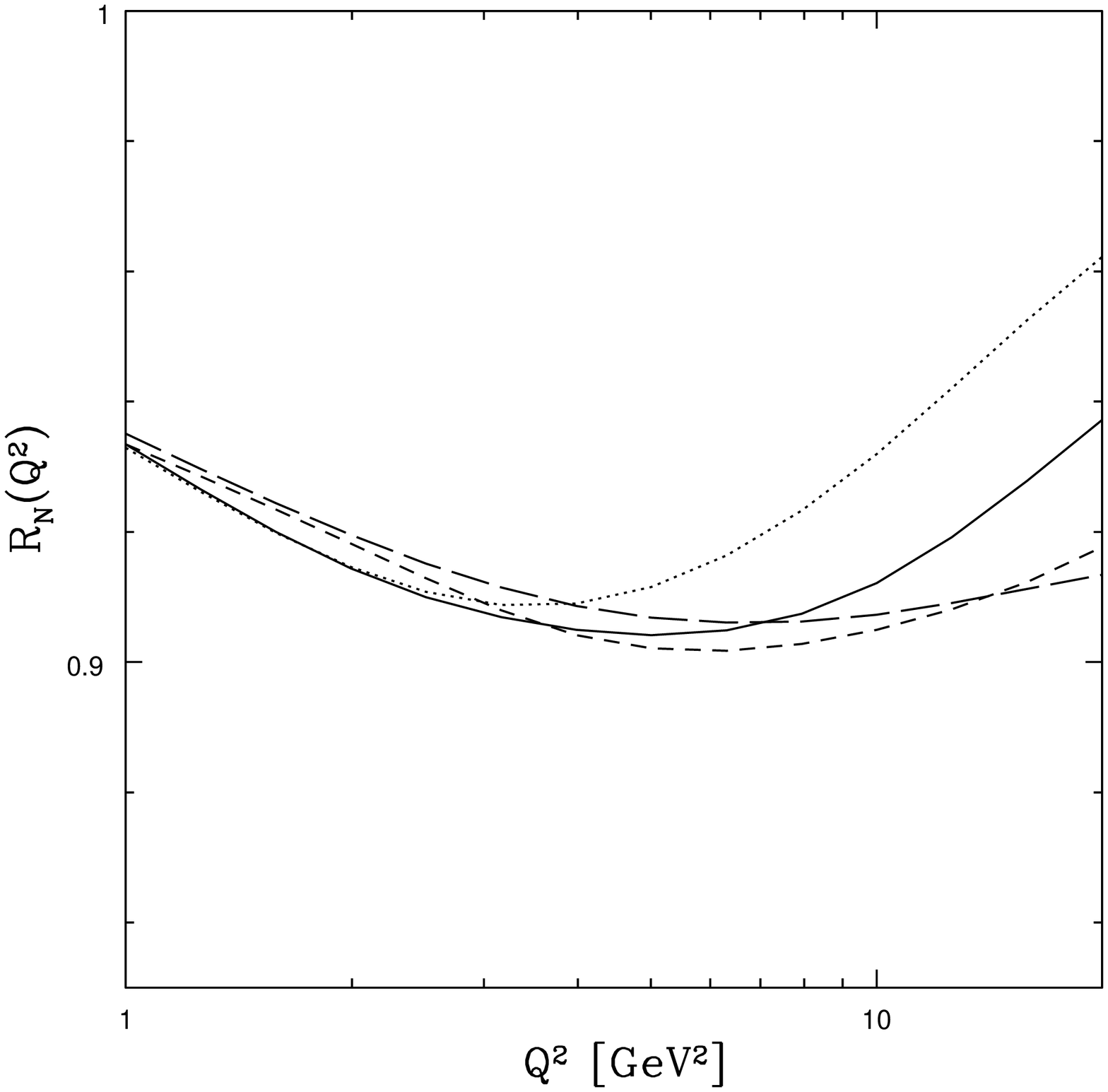}
\epsfysize=6cm \epsfbox{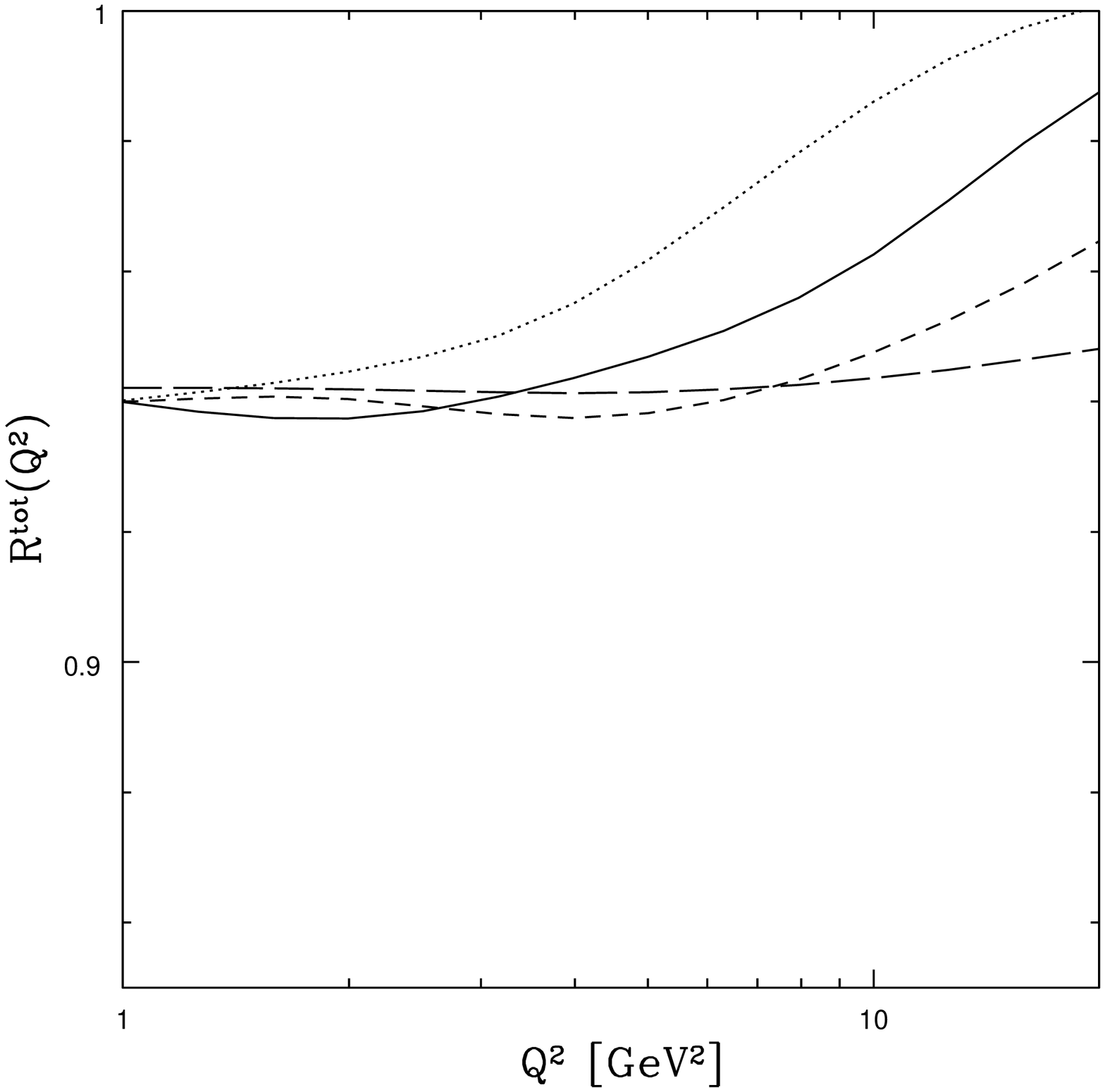}
}
\vspace{1 cm}
\caption[ ]{$R_N(Q^2)$ (eq. (17)) and of $R^{tot}(Q^2)$ (eq. (18))
at $x=1$ for the deuteron target
with different choices of the
excited nucleon mass: $m^*=1.44$ (GeV) dotted line, $m^*=1.8$ (GeV)
continuous line, $m^*=2.4$ (GeV) short-dashed line. The long-dashed line
corresponds to the standard Glauber result, where no dependence of the amplitude
on the
virtuality of the external lines is taken into account.}
\label{fft1}
\end{figure}
$F_{2}^{N/d,(1)}$ is obtained summing the two discontinuities (12) and (14)
while $F_{2}^{N/d,(1)}+F_{2}^{N^*/d,(1)}$ is evaluated by taking twice the
imaginary part of (10). Looking at the
continuous curve, corresponding to an excitation mass $m^*=1.8$ GeV,
one observes that the threshold at which the FSI
starts to vanish is practically the same in $R_N(Q^2)$ and in $R^{tot}(Q^2)$,
the effect of FSI being sizably smaller in the latter quantity.
 The above formalism has been extended[4] to the so called cumulative
 region, i.e. at $x>1$, where a calculation within the Schroedinger and
 Glauber approaches shows that the latter might be inadequate [5].

\section{ Transparency and momentum distributions in A(e,e'p)X processes}

At relatively low $Q^2$ FSI effects are therefore well described by
the Glauber approach. In this regime we have calculated the nuclear transparency and the
distorted momentum distributions for $^4He$ [6] and complex nuclei [7]; for
the former we have used a realistic four body wave function [8] whereas for the latter
we have developed
 a number-conserving linked cluster expansion for the distorted
 one-body mixed density matrix
 with Glauber multiple scattering theory and with correlated nuclear wave
 functions containing  realistic
 central and non-central correlations[7]. The linked cluster expansion includes
  final state interactions at all orders in
the number of rescatterings while
initial state correlations are taken into account at the lowest order.
The results of the transparency
for $^{16}O$ and
$^{40}Ca$ are summarized in the tables, where $i$) $T_{SM}$ is the value of
transparency in
the case of an uncorrelated nuclear wave function,
$ii$) in $\Delta T^{SM}_{FSI}$ correlations are includes but FSI takes place with
uncorrelated nucleons only (notice that in case of no FSI this term would be
zero), $iii$) $\Delta T^{H}_{FSI}$ is the hole contribution (the struck nucleon
is correlated), $iv$) $\Delta T^{S,1}_{FSI}$ and $\Delta T^{S,2}_{FSI}$ are the
spectator contributions (FSI
takes place with a correlated nucleon). In the linked expansion, as a
consequence of the
constraint of the conservation
of the number of nucleons, one obtains in fact two spectator terms with opposite sign.

%%%%%%%%%%%%%%%%%%%%%%%%%%%%%%%%%%%%%%%%%%%   MAX
\begin{figure}[!ht]
%\vskip{2 cm}
\centerline{
\epsfysize=4.3cm \epsfbox{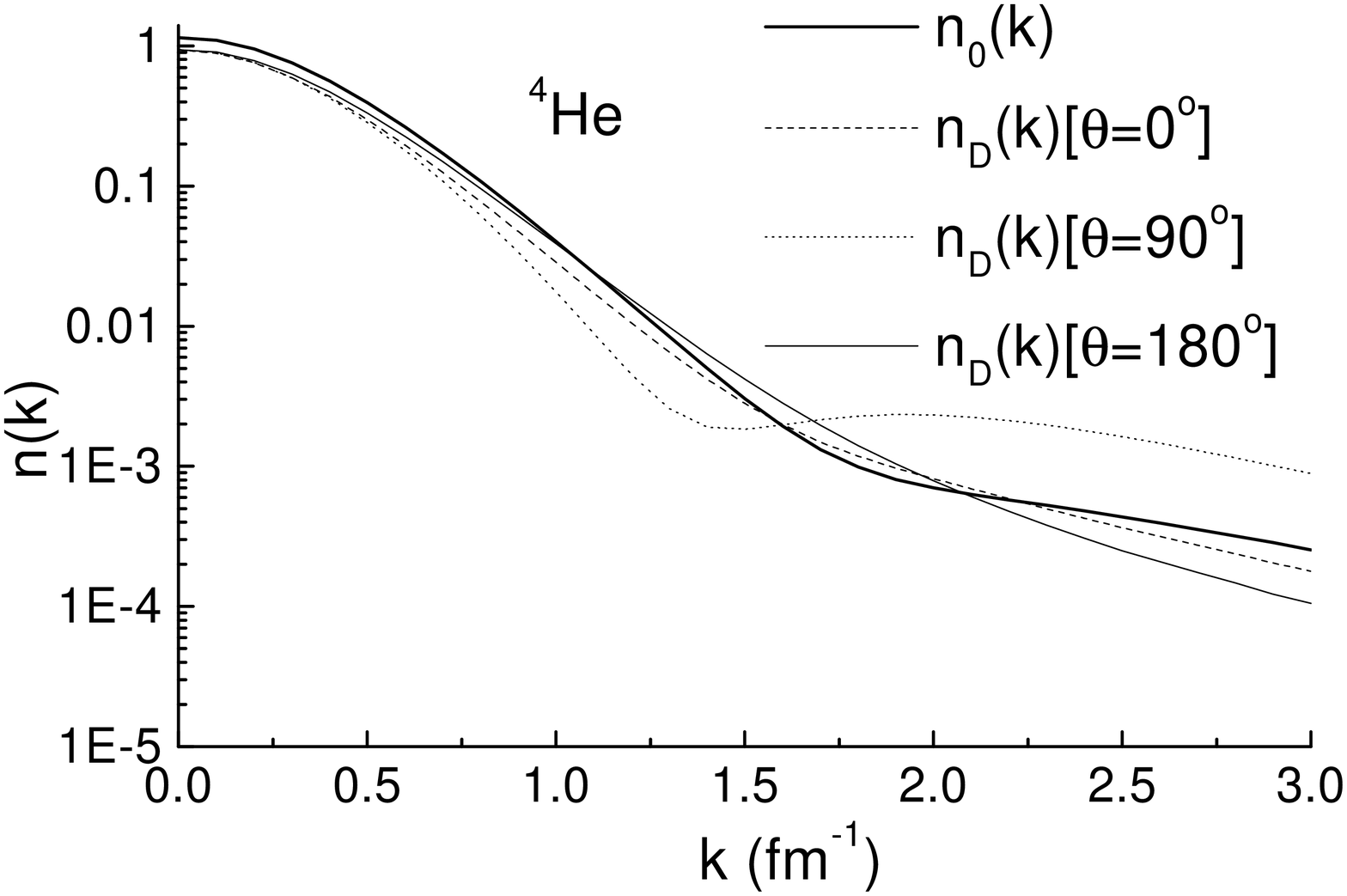}
\epsfysize=4.3cm \epsfbox{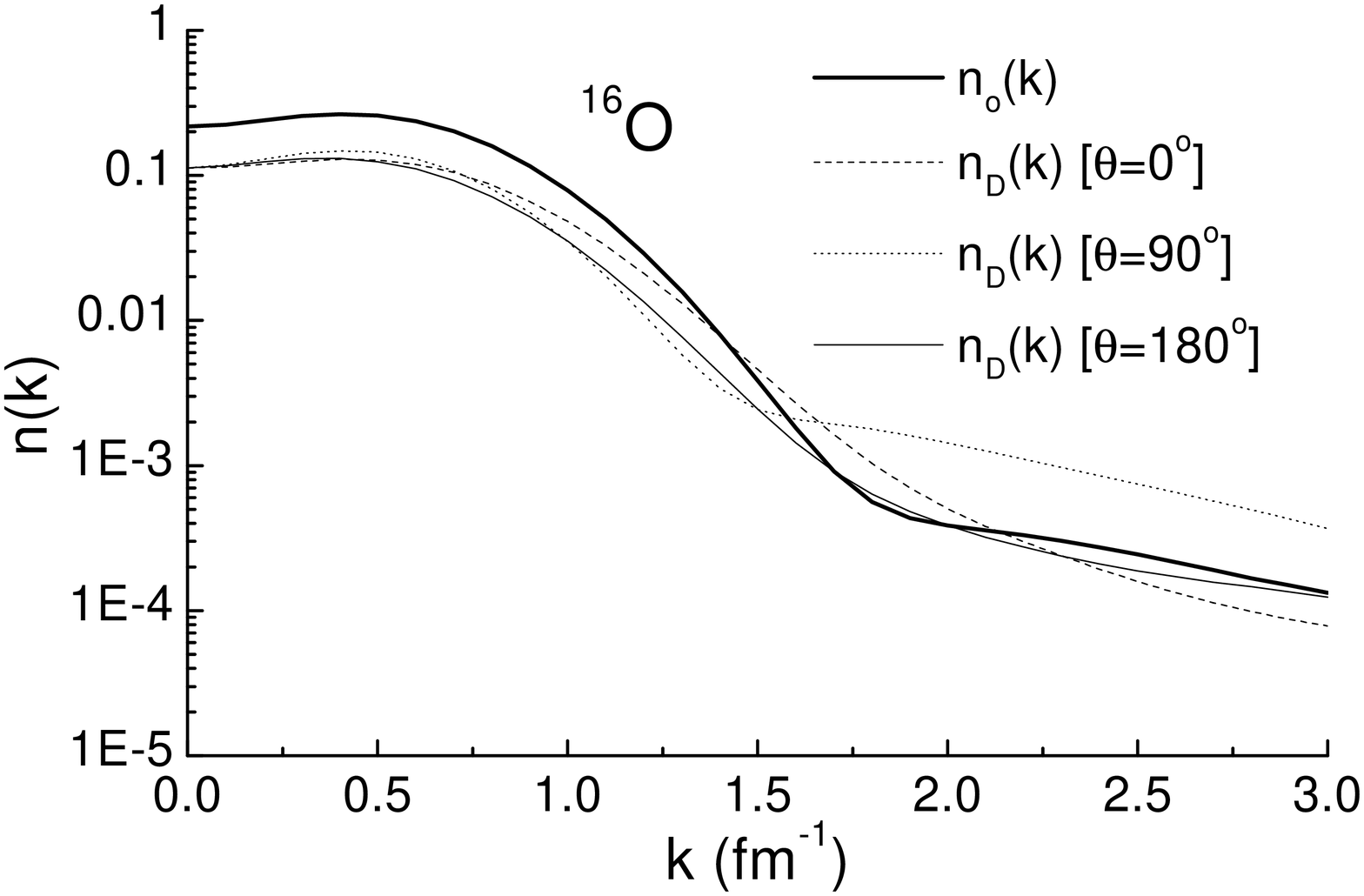}
}
%\vspace{1 cm}
\caption[ ]{Distorted momentum distribution in $^4He$ and ${ }^{16}O.$ The full line
represents the undistorted momentum distributions and the broken lines
the distorted momentum distributions at {\it parallel} $(\Theta= 0^o)$,
{\it antiparallel} $(\Theta = 180^o)$ and {\it  perpendicular} $(\Theta
= 90^o)$ kinematics.}
\label{fsi}
\end{figure}
%%%%%%%%%%%%%%%%%%%%%%%%%%%%%%%%%%%%%%%%%%%   MAX

The distorted momentum distributions for $^4He$
and $^{16}O$ are shown in Fig. 3.

\begin{table}
\caption[dummu6]{The nuclear transparency for $^{16}O$.}
\begin{flushleft}
\renewcommand{\arraystretch}{1.2}
\begin{tabular}{llllllllllll}
\hline\noalign{\smallskip}
   &   \qquad  $T_{SM}   \qquad  $  &  $\Delta {T^{SM}_{FSI}}$ \qquad &
   $\Delta {T^{H}_{FSI}}$ \qquad &  $\Delta{T^{S,1}_{FSI}}$ \qquad &
$\Delta{T^{S,2}_{FSI}}$ \qquad& $T$  \qquad \\
\hline
Central    & \qquad 0.51 & 0.020 & 0.032  & --0.013 & 0.022 & 0.57 \\
Realistic  & \qquad 0.51 & 0.003  & 0.009 & 0.001 & --0.001 & 0.52 \\
\noalign{\smallskip}\hline
\end{tabular}
\renewcommand{\arraystretch}{1}
\label{table1}
\end{flushleft}
\end{table}
\begin{table}
\caption[dummu6]{The nuclear transparency for $^{40}Ca$.}
\begin{flushleft}
\renewcommand{\arraystretch}{1.2}
\begin{tabular}{llllllllllll}
\hline\noalign{\smallskip}
   &   \qquad  $T_{SM}   \qquad  $  &  $\Delta {T^{SM}_{FSI}}$ \qquad &
   $\Delta {T^{H}_{FSI}}$ \qquad &  $\Delta{T^{S,1}_{FSI}}$ \qquad &
$\Delta{T^{S,2}_{FSI}}$ \qquad& $T$  \qquad \\
\hline
Central    & \qquad 0.41 & 0.020 & 0.028  & --0.011 & 0.023 & 0.47 \\
Realistic  & \qquad 0.41 & 0.002  & 0.008 & --0.001 & 0.001 & 0.42 \\
\noalign{\smallskip}\hline
\end{tabular}
\renewcommand{\arraystretch}{1}
\label{table1}
\end{flushleft}
\end{table}
\vspace{.3 cm}

Our findings can be summarized as follows:
 \begin{enumerate}
      \item The effect of NN correlation on nuclear transparency $T$ amounts
      to
$\sim 3\%$. The results of the exact calculation for $^4$He are
consistent with those obtained with the
cluster expansion for $^{16}O$ and $^{40}Ca$, indicating a rapid convergence of
our number conserving cluster expansion.
The small correction term is due to a cancellation between the short-range
repulsive correlation and the intermediate-range attractive correlation terms,
so that a
calculation with short-range correlations only would give misleading conclusions.
 \item The \textit{spectator} effect is very small and thus there is no
significant cancellation between the \textit{spectator} and \textit{hole}
terms.
  \item Double rescattering terms amount to $14\%$ of the leading order
term, namely the single rescattering contribution.
Thus double rescatterings cannot be neglected.
    \item As for the momentum distribution the FSI dominates the high momentum
    component in the direction
perpendicular to the direction of the virtual-photon momentum, though its magnitude is reduced if one takes into
account the tensor-type correlation which induce the D-wave component in
$^4$He.
   \item  On the contrary, in the directions parallel and anti-parallel to the
   virtual-photon direction,
   the effect of FSI is sizably smaller.
  \end{enumerate}
\vskip.15in

\section{ Final state interaction  in exclusive A(e,e'p)B processes}

The formalism described in Section 1 has also been applied to the
calculation of {\it exclusive} processes [10]. The results for the
process$^3He(e,e'p)^3H$ are presented in Figs. 4 and 5, where the effects from Glauber
 rescattering and Finite Formation Time  are displayed for various values
 of $Q^2$.  It can be seen that  Glauber Final State Interactions has a large effect,
in that it completely washes out the diffraction  dip predicted by the
Plane Wave Impulse Approximation; on the other hand side, they exhibit
a very mild $Q^2$ dependence, unlike  Finite Formation Time effects which strongly depend
 upon $Q^2$, in such a way that at  $Q^2 \simeq 20
{GeV}^2$  the Plane Wave Impulse Approximation  result is almost completely recovered.
%%%%%%%%%%%%%%%%%%%%%%%%%%%%%%%%%%%%%%%%%%%%%%%%%%%%%%%%%%%%%%%%%%%%%%%%%%%%%
\begin{figure}[!ht]
%\vskip 8cm
\centerline{
\epsfysize=7cm \epsfbox{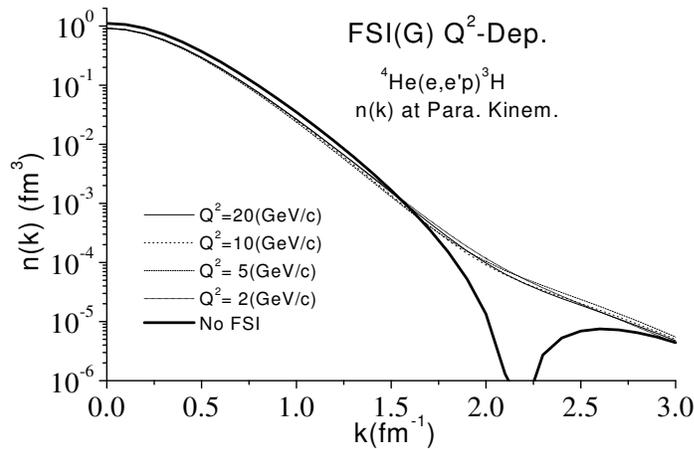}
%\special{psfile=FigDue.ps hoffset=170 voffset=0 hscale=40 vscale=40 angle=90}
}
%\vskip -.5cm
\caption[ ]{The effects from Glauber rescattering
      on the  distorted momentum distributions in the
      exclusive  process $^4He(e,e'p)^3H$ at parallel kinematics. The full curve represents
      the Plane Wave Impulse Approximation result, whereas the broken lines include the
      Glauber rescattering at various values of $Q^2$}
\label{parakinem1}
\end{figure}
\newpage
\begin{figure}[!ht]
%\vskip 8cm
\centerline{
\epsfysize=7cm \epsfbox{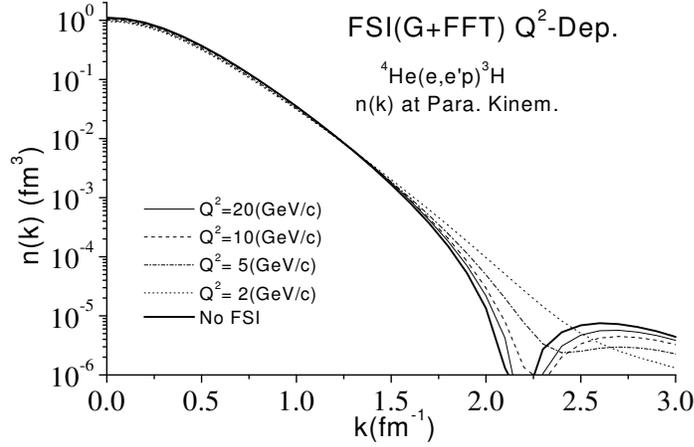}
%\special{psfile=FigUno.ps hoffset=170 voffset=0 hscale=40 vscale=40 angle=90}
}
%\vskip -.5cm
\caption[ ]{Same as   Figure 4,  with Glauber rescattering plus Finite
      Formation Time effects.}
\label{parakinem2}
\end{figure}
%%%%%%%%%%%%%%%%%%%%%%%%%%%%%%%%%%%%%%%%%%%%%%%%%%%%%%%%%%%%%%%%%%%%%%%%%%%%%

\textbf{References}
\vskip.15in

\noi [1] V.N.Gribov, {\it Sov. Phys. JETP} {\bf 29}, 483 (1969);
{\bf 30} (1970) 709; L. Bertocchi, {\it Nuovo Cimento} {\bf 11A}, 45 (1972);

\noi [2] M.A.Braun, C.Ciofi degli Atti and D. Treleani {\it Phys. Rev.} {\bf
C62}, 034606 (2000);

\noi [3] L. Frankfurt, W.R. Greenberg, G.A. Miller and M.
Strikman, {\it Phys. Rev.} {\bf C46}, 2547 (1992);

\noi [4] M. Braun, C. Ciofi degli Atti, L. Kaptari and D. Treleani, to
appear;

\noi [5] C. Ciofi degli Atti, L. Kaptari, and D. Treleani, {\it Phys. Rev.},
to appear;

\noi [6] H. Morita, C. Ciofi degli Atti  and D. Treleani, {\it Phys. Rev.},
{\bf
C60}, 034603 (1999);

\noi [7] C. Ciofi degli Atti  and D. Treleani, {\it Phys. Rev.},
{\bf C60}, 024602 (1999);

\noi  [8] M. Alvioli,  C. Ciofi degli Atti H. Morita  and D. Treleani,
to appear;

\noi [9] H.Morita, Y.Akaishi, O.Endo and
H.Tanaka {\it Prog.Theor. Phys.} {\bf 78}, 1117 (1987);

\noi [10] M. Braun, C. Ciofi degli Atti, H. Morita and D. Treleani, to
appear.

\end{document}